\documentclass[lettersize,journal]{IEEEtran}
\usepackage{amsmath,amsfonts}
\usepackage{algorithmic}
\usepackage{algorithm}
\usepackage{array}
\usepackage[caption=false,font=normalsize,labelfont=sf,textfont=sf]{subfig}
\usepackage{textcomp}
\usepackage{stfloats}
\usepackage{url}
\usepackage{verbatim}
\usepackage{graphicx}
\usepackage{cite}
\usepackage{makecell}
\usepackage{multirow}
\usepackage{booktabs}
\usepackage{soul}
\usepackage[dvipsnames]{xcolor}

\usepackage{authblk}
\usepackage{color}

\begin{document}
\title{\vspace{-0.5cm} Practical Implications of Implementing Local Differential Privacy for Smart grids}

\author[1]{Khadija Hafeez}
\author[1]{Mubashir Husain Rehmani}
\author[2]{Sumita Mishra}
\author[1]{Donna O'Shea}
\affil[1]{Munster Technological University (MTU), Cork, Ireland }
\affil[2]{Rochester Institute of Technology (RIT), Rochester, NY, USA }



\maketitle

\begin{abstract}

Recent smart grid advancements enable near-real-time reporting of electricity consumption, raising concerns about consumer privacy. Differential privacy (DP) has emerged as a viable privacy solution, where a calculated amount of noise is added to the data by a trusted third party, or individual users perturb their information locally, and only send the randomized data to an aggregator for analysis safeguarding user’s and aggregator’s privacy. However, the practical implementation of a Local DP-based (LDP) privacy model for smart grids has its own challenges. In this paper, we discuss the challenges of implementing an LDP-based model for smart grids. We compare existing LDP mechanisms in smart grids for privacy preservation of numerical data and discuss different methods for selecting privacy parameters in the existing literature, their limitations and the non-existence of an optimal method for selecting the privacy parameters. We also discuss the challenges of translating theoretical models of LDP into a practical setting for smart grids for different utility functions, the impact of the size of data set on privacy and accuracy, and vulnerability of LDP-based smart grids to manipulation attacks. Finally, we discuss future directions in research for better practical applications in LDP-based models for smart grids.

\end{abstract}

\vspace{-0.5cm}
\section{Introduction}

Smart meters, equipped with bidirectional communication capabilities, are integral components of smart grids. Fine-grained power consumption measurements or metered data of a consumer are reported to the utility provider at frequent intervals, typically every $10-15$ minutes, for efficient demand forecasting, electricity production and distribution. However, these highly granular measurements can pose serious privacy threats to electricity consumers, as sensitive information such as consumer behavior, lifestyle, and financial status can be inferred from this data. Further, in order to mitigate risks of customers suing the power utility companies for mishandling sensitive data and to comply to region's privacy polices, electric utilities should implement robust privacy policies using privacy enhancing technologies that align with both state and federal privacy laws and regulations of the region, thereby reducing their exposure to liability.
To address these privacy concerns, various privacy preserving techniques have been proposed, among which Differential privacy (DP) has emerged as a prominent technique for preserving user privacy in smart grid environments due to its capability of providing a formal mathematical proof for privacy guarantees, scalability and efficiency as compared to other storage and compute intensive privacy enhancing techniques, such as federated learning \cite{deSAMP_fedDP} and homomorphic encryption \cite{dream}. In addition, local differential privacy (LDP) with decentralized mechanisms eliminates the need for a trusted third party, unlike anonymization methods that rely on a central escrow as an anonymizer.

DP is a data obfuscation technique which leverages a probabilistic model for privacy preservation in smart grids by adding a highly curated noise to the data by carefully selecting privacy parameters such as sensitivity and privacy budget or epsilon ($\epsilon$). 
In Central DP setting noise added by a trusted third party whereas in a local DP (LDP) setting, a differentially private noise is directly added to the metered data by the smart meter, before sending it to the aggregator. This enables a more distributed deployment of the privacy-preserving mechanism for the energy consumption data.\par

Depending on the data type (categorical or numeric), different LDP mechanisms to curate the noise that provides privacy guarantees have been proposed in the literature, including Randomize Response, Unary Encode, Laplace mechanism and more. Different variations of these mechanisms have been used for multiple applications, however, in the context of the smart grids, the options are limited. Since the metered data is numeric, the most widely used option is using the Laplace mechanism, where each smart meter generates a random variable from Laplace distribution and adds to the original meter data. Another option for numeric data in LDP is Duchi's realization of Randomized Response (RR)\cite{duchi2013local}. However, this mechanism has not been extensively explored in the context of smart grids due to the lack of mechanisms to cater to large numeric input domain sizes and increased errors due to data discretization.\par

In this article, we discuss the challenges and practical implications of implementing LDP mechanisms for numeric time series metered data in smart grids. These challenges include the selection of privacy parameters, i.e., sensitivity and epsilon, and its impact on the accuracy of aggregation results; impact of aggregation group size on the privacy guarantee; the challenge of privacy loss due to continual data release; and the vulnerability of LDP-based models for smart grids against manipulation attacks.

To date, to the best of our knowledge, the impact of these challenges within the context of smart grid systems has not been previously explored. Given this gap, in this article we will discuss these challenges and provide future direction and guidance for researchers leveraging LDP-based models for smart grids.

Our contributions in this article include the following; 
\begin{itemize}
    \item Comparative analysis of existing literature on LDP-based privacy models for smart grids. 

    \item Discussion on challenges of translating and adapting theoretical LDP for practical applications in smart grids.
    \item Future directions for comprehensive LDP-based model design in smart grids. 
\end{itemize}
The rest of the article is organised as follows. Relevant LDP mechanisms are discussed in section \ref{LDP_mech}, followed by existing literature in LDP-based models for smart grids in section \ref{Existing_LDP_SG}. The limitations and challenges of implementing LDP-based models for smart grids are discussed in section \ref{Limitations}. Finally, the future directions for practical adaptation of LDP-based models for smart grids are presented in section \ref{Future}.

\section{Preliminary Knowledge}\label{LDP_mech}
In this background section, a brief overview of the operational principles of LDP mechanisms, specifically Laplace and Randomized Response methodologies, is provided. 
\vspace{-0.4cm}


 \subsection{Laplace mechanism} 

The Laplace mechanism for LDP adds noise to the individual data points and the privatized data is then published. Each user’s individual data point is perturbed by adding noise, which is a random variable drawn from a Laplace distribution with the probability density function taking mean and scale as parameters. The scale parameter is directly proportional to the sensitivity and inversely proportional to the privacy budget. Sensitivity and privacy budget are the two privacy parameters generally used to define the level of privacy of a DP mechanism, as they are directly responsible for the level of noise added to the data. Due to their pivotal role in DP, setting optimal values for these parameters is imperative for Laplace mechanism to ensure privacy guarantee. \par
    
Sensitivity quantifies the influence of an individual data point on the outcome of a query, reflecting how a modification in a single input value affects the resulting output. For simplicity, we assume that each user owns a single data point which lies in the range form $-1$ to $1$ and follows Laplace mechanism to privatize data before sharing. The maximum impact of a single data point on the output of mean estimation of such data is $2$, which is the sensitivity of the LDP function.

Epsilon is the measure of privacy loss in a DP  mechanism where setting epsilon close to $0$ means highest level of privacy but it also implies maximum error due to excessive noise.
The privacy budget is chosen by the data owner as a trade-off between privacy and the utility of the data. 

\vspace{-0.4cm}
 \subsection{Randomized Response (RR)}
RR is commonly used as an LDP mechanism for categorical data. Duchi $\emph{et al.}$ \cite{duchi2013local} proposed an LDP mechanism for numerical data by conceptualizing RR, where only two possible outputs are determined by the flip of a coin.
The first step of the RR mechanism is that each data point, ranges from  $-1$ to $1$, is discretized to a set of constant values, where the chosen discrete value depends on original data point and epsilon. The discrete value is then reported using RR, where the probability of choosing perturbed value depends on epsilon.

\section{Comparative Analysis of Existing LDP Solutions for Smart Grids }\label{Existing_LDP_SG}  
In this section, the adaptation of theoretical LDP-based models for smart grids in the existing literature are analyzed, based on the methodologies for privacy parameter selection and utility functions.
\vspace{-0.4cm}
\subsection{Laplace as LDP for SG}
Laplace is the most common mechanism used in the context of privacy protection for smart grids due to its infinite divisibility property. As discussed earlier, sensitivity and privacy budget are two important privacy parameters that need to be set for LDP privacy guarantee. In the context of smart grids, depending on the utility of the energy consumption data such as billing, load monitoring, and load forecasting, existing LDP-based models choose different methods for sensitivity selection.  \par
The most common method of selecting sensitivity value for LDP-based models for smart grids is choosing the maximum value in the data set\cite{LapSGLDP,Won2016,dream}. For instance, in the case of load monitoring, where the utility function is a sum query over the electricity consumption data of all households at an instant, the sensitivity value is the measured reading of the household with maximum energy consumption at that instant. An alternative method for sensitivity selection involves utilizing the maximum variance of appliances with the highest wattage, as proposed  by Barbosa \emph{et al.} \cite{Barbosa2016}. Ebil \emph{et al.} \cite{eibl2018influence} compare choosing a global maximum and local sensitivity selection. All of the above the different sensitivity selection mechanism deviate from the theoretical sensitivity parameter selection criteria and have their own limitations. The sensitivity parameter as max overall value adds more noise than max appliance wattage, however, the latter requires more number of participants to provide privacy. The challenges of selecting an optimal sensitivity value are discussed in section \ref{Limitations}.\par
The privacy budget or epsilon is often set as $1$ in existing literature for LDP-based models for smart grids, without any particular reason. Some research \cite{eibl2018influence,SGEnv,SpectA} has been done to explore the impact of varying values of epsilon, however, no conclusive results have been drawn in determining the optimal value of epsilon.  

\vspace{-0.4cm}
\subsection{RR mechanism as LDP for SG}
In the context of smart grids, RR has been explored in a limited capacity. The only RR-based LDP model for smart grid is proposed by Gai \emph{et al.} \cite{RR_SG} which uses a discretization process to transform original continuous electricity data profile to discrete data. The process is explained below: \par 
Consider a scenario where original data point is the energy consumed in a small time interval, as measured by a smart meter, ranging from a minimum value of $0$ kWatt to a set maximum value.
The first step is to discretize data into subcategories by dividing the range into smaller sub-intervals. Now, each data point can be represented by one of the subcategories. RR is used next to report the perturbed value. At the aggregator end, frequency estimation of each discrete value can be done to get the estimated sum. 
The accuracy of aggregation in the RR mechanism depends on several factors, including the number of sub-intervals, the number of participants, epsilon, and the range of the input data domain. However, the model proposed by Gai \emph{et al.} lacks practical comparison with Laplace-based LDP mechanisms concerning both accuracy and privacy.

\section{Challenges in Applying Theoretical LDP-based Models to Smart Grids Applications }\label{Limitations} 
In this section, the challenges and limitations of translating a theoretical LDP-based model for practical data privacy of smart grids are discussed. 
\vspace{-0.4cm}
\subsection{Lower Data Utility}
The challenge of using Laplace as an LDP mechanism for smart grids lies in calibrating the noise as Laplace samples noise that can range from $-\infty$ to $\infty$. Due to unbounded sample range of the Laplace mechanism, the output of perturbed data can lie in the range from $-\infty$ to $\infty$. In Laplace-based LDP, given that it is possible to draw noise from a negative range, an energy consumer can report a negative value as amount of energy consumed, even though in reality, the measure of energy consumption cannot be negative. The unbounded sample range also means when employing a smaller epsilon value to increase privacy, the resulting higher noise levels can greatly reduce the utility of the data. 
\begin{figure}
    \centering
    \includegraphics[width=1\linewidth]{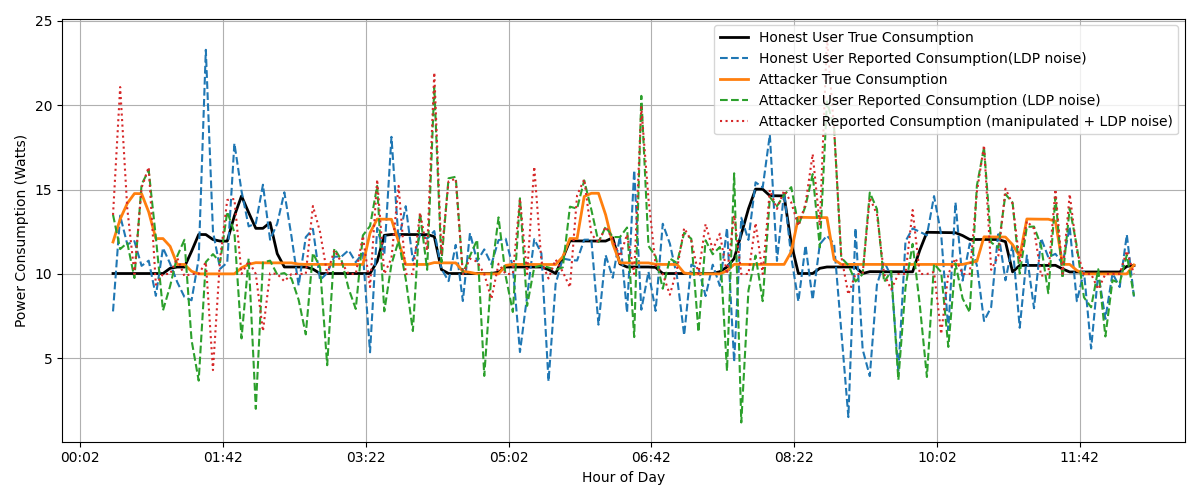}
    \caption{Vulnerability of LDP mechanism to manipulation attack as attacker can not be identified due to noise.}
    \label{fig:manipulationattack}
    \vspace{-0.5cm}
\end{figure}
\vspace{-0.4cm}
\subsection{Vulnerability to Manipulation attacks}

LDP safeguards individual privacy by adding noise to data at the source, but this can obscure accurate consumption details, complicating detection of intentional misreporting. Hence, the LDP architecture by design is vulnerable to manipulation attacks and can be exploited by adversaries. For instance, a malicious smart meter may aim to lower its utility bills by concealing its true data behind the LDP noise, thereby reporting manipulated information (shown in Fig. \ref{fig:manipulationattack}).
Thus, legal frameworks must clearly define the roles of energy providers and consumers while addressing LDP's technical complexities to balance privacy and compliance.
Furthermore, when multiple malicious smart meters collaborate, they can orchestrate manipulation attacks to disrupt the services of a power grid. As explored by Wang \emph{et al.} \cite{prefLimitationLap} Laplace-based LDP model is found to be more vulnerable to manipulation attacks due to the unbounded noise addition, resulting in maximum possible risk of adversarial attack.


\subsection{Challenge of setting the Sensitivity Parameter}

Setting an optimal criteria for selecting sensitivity values in accordance with the theoretical model of LDP presents a significant challenge, primarily because the sensitivity parameter depends on the nature of the input data and query functions. The difficulty in determining the appropriate sensitivity value also arises from the need to translate theoretical sensitivity values into practical applications, as the theoretical sensitivity values often result in high levels of noise, rendering the data practically unusable. For example, according to the theoretical LDP model, the sensitivity value for spatio-temporal aggregation in smart grids is defined as the sum of the maximum energy consumption by any single household within an instant over a billing period. However, in practice, this theoretical sensitivity parameter is not selected due to the excessive noise it introduces, rendering the data ineffective. \par

Given the above challenges, the sensitivity value is often set to the maximum energy consumption of a single household over a given time period rather than following the theoretical sensitivity value \cite{dream,LapSGLDP,Won2016}.
However, this approach encounters practical challenges, as individual consumers typically lack knowledge of the maximum energy consumption levels of other smart meters. Even if such information were assumed to be available, the highest energy consumption value may vary within each interval, consequently altering the sensitivity value at each instance. This variability of sensitivity value at each instant compromises privacy guarantees. One way to address this issue is to preset the range of energy consumption in an instance \cite{eibl2018influence}.
The minimum power consumption by an individual consumer in $10$ minutes can be set to $0$ and the maximum can be set to 42kWatt, as per three phase household in Germany. However, analysis of pre-setting an upper bound as the sensitivity value shows that it causes higher levels of inaccuracy, as it is highly unlikely that a household reaches the upper bound limit of power consumption often. Most individual consumers do not consume the maximum possible energy and this results in more noise with little to none utility for individual data release. Further, this approach fails in cases where there are households with renewable energy resources that can record negative energy consumption to show net-metering while supplying energy back to the grid.  \par
Alternatively Barbosa \emph{et al.} uses the appliance with the highest wattage as the sensitivity value in their model. However, this method offers limited protection against privacy attacks, such as occupancy detection and individual membership inference attacks. Additionally, it is not well-suited for load monitoring, as it requires a considerable number of participants for accurate aggregation results. Furthermore, this approach is closely linked to user behavior, where users with lower energy consumption experience higher levels of noise in the data.

\begin{figure}
    \centering
    \includegraphics[width=0.8\linewidth]{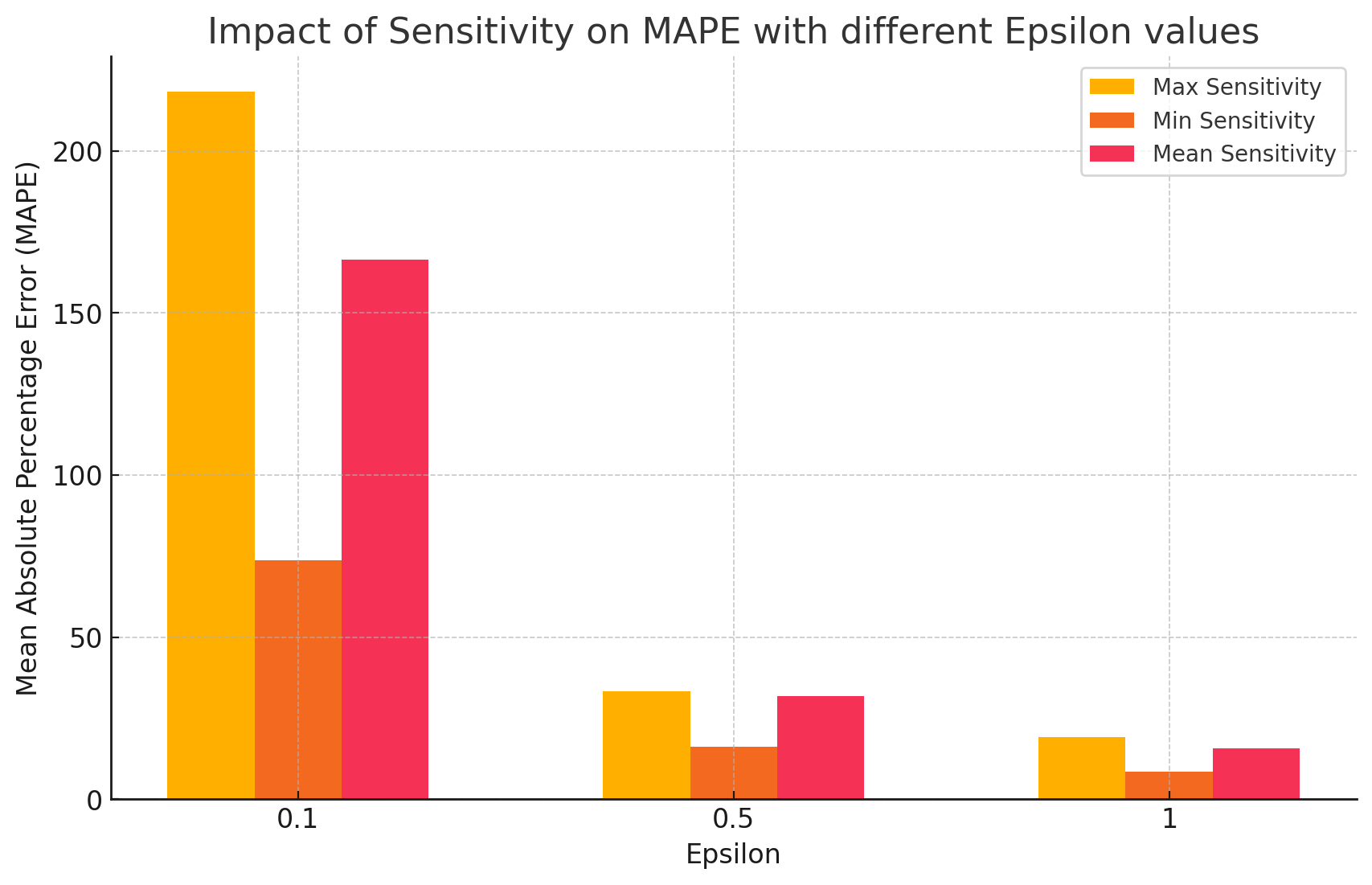}
    \caption{Impact of selecting different privacy parameters on accuracy.}
    \label{fig:sen_comp}
    \vspace{-0.5 cm}
\end{figure}

Different methods of selecting privacy parameters in existing literature are compared in Table \ref{tab:lit}. It can be seen that a theoretical sensitivity value is not used for LDP-based models for smart grids as it introduces a lot of noise and reduces the utility of the data significantly. As seen in Fig. \ref{fig:sen_comp}, increasing the sensitivity value leads to a corresponding increase in the noise level. Moreover, compromised sensitivity values further diminish the privacy guarantee and complicate the analysis of privacy loss.
\vspace{-0.3 cm}
\subsection{Optimal Epsilon value}

Selecting an optimal epsilon value remains an open challenge in the field of DP, as it determines the level of privacy provided by the mechanism. In the context of smart grids, one of the reasons for the issue of selecting a suitable epsilon value is the lack of a standardized performance metrics to evaluate the privacy model. Various values of epsilon have been employed in existing literature, ranging from $0.1$ to $6$, as can be seen from Table \ref{tab:lit}. However, the extent to which privacy is preserved in practical scenarios remains largely unexplored. For instance, while setting epsilon to $1$ theoretically implies no privacy, many studies employing Laplace as LDP have used this value in experimental analyses of their models.  Ebil \emph{et al.} \cite{eibl2018influence} experiment with a range of epsilon values for load forecasting, however, the model only provides membership privacy, i.e., the membership of an individual household from aggregate results is protected. It does not provide protection to individual household load profile from Non Intrusive Load Monitoring (NILM) attacks or occupancy detection attacks. 

\begin{table*}[t]
  \centering
  \caption{Comparison of LDP-based models for smart grids }
  \begin{tabular}{|m{0.02\textwidth}|m{0.06\textwidth}|m{0.2\textwidth}|m{0.08\textwidth}|m{0.1\textwidth}|m{0.4\textwidth}|}
  \hline
  \thead{Ref.} & \thead{LDP Mech.} & \thead{Sensitivity ($\Delta f$)}  & \thead{Privacy ($\epsilon$)} & \thead{Utility Function}& \thead{Limitations in Privacy Analysis}\\
  \hline
   \cite{LapSGLDP} & Laplace & Theoretical $\Delta f$: Sum of max values by each consumer. Experimental $\Delta f$: Maximum overall value in the data set
 & $\epsilon = 1$ & Load Monitoring & \begin{itemize} 
   \item The experimental privacy parameters differ from theoretical analysis.
   \item For continuous queries in successive intervals for a day requires $\epsilon$ = $23$. 
   \item LDP output is combined with random permutation of data to achieve privacy guarantee. 
   \end{itemize}
\\ \hline
    \cite{Won2016} & Laplace & Theoretical $\Delta f$: Sum of power demand of all appliance.  
Experimental $\Delta f$: max power consumption of individual consumer
 & $\epsilon = 1$ & Load Monitoring & \begin{itemize}
        \item The model provides $(\alpha-\epsilon)$DP protection.
        \item No experimental analysis of privacy loss with global vs. individual sensitivity is done.
        \item The choice of setting $\epsilon = 1$ for experiments as appose to the theoretical model is unclear.
    \end{itemize}  
\\ \hline
    \cite{dream} & Laplace & Theoretical $\Delta f$: Sum of max values by each consumer.
Experimental $\Delta f$: Maximum overall value in the data set
 & $\epsilon = 1$  & Load Monitoring & \begin{itemize}
        \item The model provides privacy for sum of measurements, not the individual profile.
        \item Used encryption on top of LDP for privacy guarantee. 
    \end{itemize}
\\ \hline
    \cite{eibl2018influence} & Laplace &  Theoretical $\Delta f$: maximum power demand fused in a residential homes. 
 Experimental $\Delta f$: highest power demand per hour in the whole data set, and its $90th$, $99th$ percentiles. 
 &  $\epsilon = 0.5 - 0.9 $ & Electricity demand forecast & \begin{itemize}
        \item Individual membership in aggregated sum is protected. 
        \item The individual data profile as well as individual aggregated sum is not protected. 
        \item The privacy loss increases with gradual data release.

        \end{itemize} \\ \hline
    \cite{edpnct} & Laplace&   $\Delta f$: Max, Mean, Min of the overall dataset. 
 &  $\epsilon = 1 $ & Load Monitoring and Billing & 
 \begin{itemize}
        \item Vulnerable to manipulation attacks 
        \item Communication overhead.

        \end{itemize} \\ \hline
    \cite{Barbosa2016} & Laplace & $\Delta f$: Maximum variation of the electrical appliance with highest wattage  & $\epsilon = 1.07$ & Load Monitoring, Billing & \begin{itemize}
        \item Appliance membership in a load profile is protected.  
        \item Requires a very large data set to provide appliance privacy. 
        \item Individual membership privacy in aggregated load is not discussed.  
    \end{itemize}
\\ \hline
    \cite{SGEnv} & Laplace & 1 & $\epsilon = 0.01-0.5 $ & Billing & \begin{itemize}
        \item No explanation of how sensitivity is set.
        \item The privacy loss analysis is not done. 
    \end{itemize}  \\ \hline
    \cite{SpectA} & Laplace  & $\Delta f$: Maximum difference between any two data record for all appliances & $\epsilon = 0.1-2$ & Error in total energy consumption. & \begin{itemize}
        \item Data obfuscation done on appliance level is practically in feasible. 
        \item Appliance classification in a aggregated load profile is protected but individual smart meter membership in aggregated load monitoring is not protected under the privacy model.

    \end{itemize}\\ \hline
    \cite{RR_SG} & RR  & N/A & $\epsilon = 1-6$ &  N/A & \begin{itemize}
        \item No accuracy analysis is done for real smart meter data. 
        \item Privacy analysis does not reflect the level of privacy protection. 
    \end{itemize}\\ \hline
   
  \end{tabular} 
  \label{tab:lit}
\end{table*}


\subsection{Impact of Aggregation Group Size}
\begin{figure}[!t]
    \centering
    \includegraphics[width=1\linewidth]{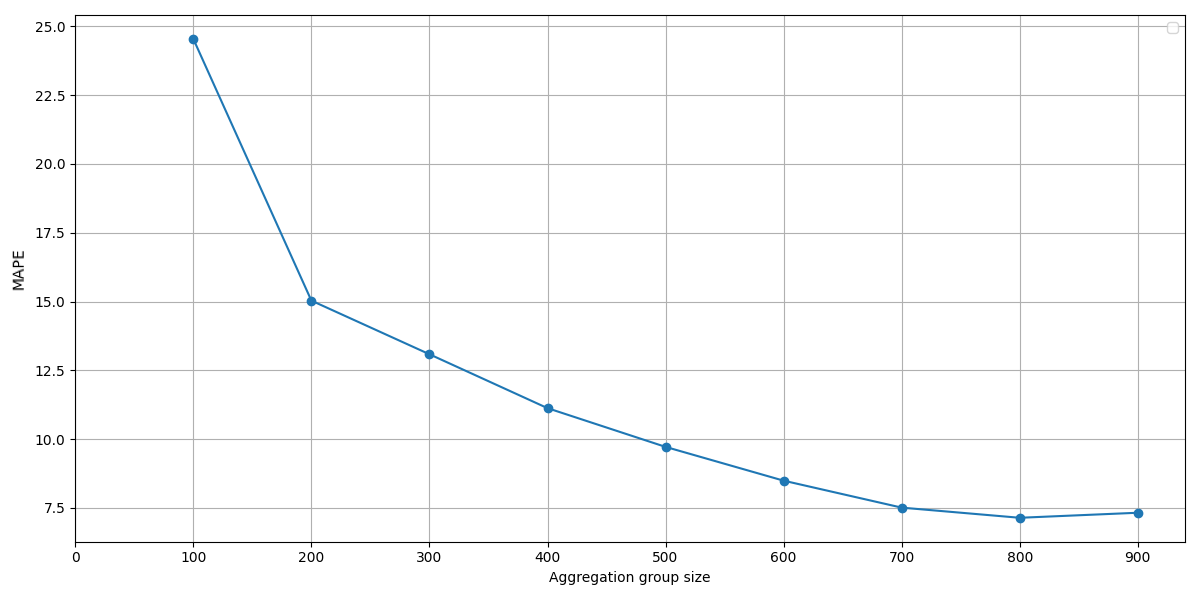}
    \caption{Impact of aggregation group size on MAPE.}
    \label{fig:aggrGsize}
    \vspace{-0.4 cm}
\end{figure}

 The dependence of LDP-based smart grids on the size of the aggregation group presents a significant challenge, primarily due to the difficulty in meeting the minimum aggregation size requirements in practice. This minimum aggregation group size is crucial for the effectiveness of LDP-based smart grids, as LDP relies on the number of participants or data points. As the number of participants increases, the impact of the noise added by LDP decreases. Here, the size of the aggregation group can be the number individual consumers in an area reporting individual metered data for load monitoring, or it can be the number of metered data reported in a billing period by an individual. Hence, a small number of smart meters participating in LDP-based smart grid for load monitoring is not suitable as the error in results increase with the fewer participants in aggregation as depicted by Fig. \ref{fig:aggrGsize}. 
\subsection{Continual Privacy Loss in Time Series Data}
Continual privacy loss in the time series data is a major concern for practical development of LDP-based privacy models for smart grids, mostly due to the limited privacy budget and the composition property of DP. Most existing literature works overlook the impact of continuous release of private data and treat data as a static time snapshot when analysing their LDP-based model for smart grids. However, in a practical LDP setting, the individual consumer releases private smart meter data in each time slot. As discussed by Ebil \emph{et al.} \cite{eibl2018influence}, the continual release of such time series data consumes privacy budget over time. Consequently, after a certain time period, the LDP data will no longer provide privacy guarantees and will begin to leak information.

\subsection{Limitations of RR}
\begin{figure}
    \centering
    \includegraphics[width=1\linewidth]{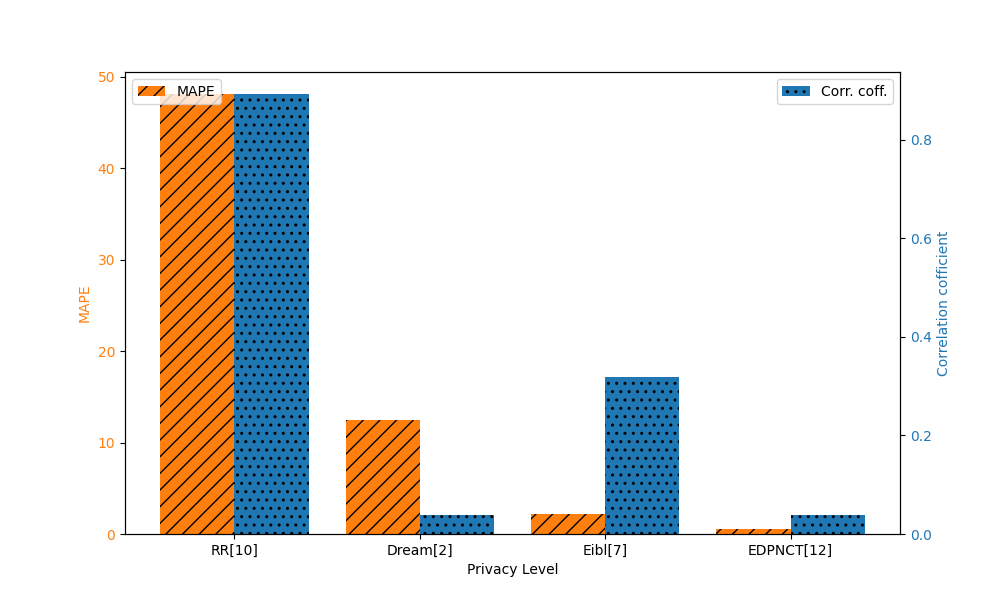}
    \caption{Comparison of different LDP models for smart grids. }
    \label{fig:privlevel}
    \vspace{-0.5cm}
\end{figure}

There is limited work in this category, but the primary challenge of employing RR as an LDP mechanism for the smart grids lies in the lack of an optimal discretization method, as it depends on the privacy parameters and the original distributions of the numeric input data. Discretizing data requires knowledge of the input data domain and the number of subcategories may also impact the privacy guarantee and the accuracy of the estimated sum. This limitation is highlighted in the sole existing work utilizing RR for smart grids by Gai et al. \cite{RR_SG}.  As seen in Fig. \ref{fig:privlevel}, Laplace mechanism employed in EDPNCT\cite{edpnct}, Dream \cite{dream}, and Eibl \cite{eibl2018influence} performs better than mechanism RR \cite{RR_SG} due to lack of an optimal discretization method.\par
Another obstacle in using RR as a privacy mechanism for smart grids is its inefficient performance in term of accuracy for epsilon less or equal to $2 $. As discussed by Wang \emph{et al.} \cite{LapvsPWvsH}, the inaccuracy in results for RR is still high at higher epsilon values, since RR has only two possible outputs. Since the input can take any values ranges from $-1$ to $1$, the output should have many possibilities for higher epsilon values, which results in lower privacy and high inaccuracy.

\section{Future Direction}\label{Future} 
To solve the aforementioned challenges, some research directions are identified and discussed in this section. 
\subsection{Optimal and Private determination of the Sensitivity Parameter}
There is a need for a strategy to determine and share an optimum value for sensitivity. In most existing literature works, it is assumed that each smart meter knows the maximum energy consumption in an area at each instant, in order to select the sensitivity parameter for the LDP model. However, privately sharing sensitivity has not been discussed in the literature. Since the sensitivity parameter is selected as the maximum consumed power consumption, entities consuming significantly more energy than others in the area can highly impact the outcome of the LDP model. Another research direction could be to minimize the impact of outliers to optimize the sensitivity selection.  \par 
An upper bound determined by the service provider, the highest electrical power consumption a household could have, has also been explored as the sensitivity parameter in the literature\cite{eibl2018influence}. However, utilizing this upper bound as the sensitivity parameter may yield sub-optimal accuracy in results. The sensitivity value chosen in practical experiments in existing literature \cite{dream,SGEnv,Won2016}  often differs from the theoretical LDP model, as shown in Table \ref{tab:lit}. Thus, it can be observed that the privacy guarantee does not follow the theoretical LDP model. A robust optimum sensitivity value calculation mechanism not only improves the accuracy in results, but it also decreases the privacy loss.
\vspace{-0.4cm}
\subsection{Optimal LDP mechanism for SG}
A perfect LDP mechanism for smart grids must have strong privacy guarantees, accurate data utility, computational efficiency, adaptability, and compliance with legal and ethical standards, though the inherent trade-offs make it an ideal rather than a strictly attainable standard. Existing LDP mechanisms for numeric data have been used for mean estimation function with range normalization for input domain \cite{d-priv}. 
Given the critical nature of bill calculations, wherein consumers expect precise invoicing reflecting their actual usage, a need arises for an optimal mechanism which protects numeric data across diverse ranges and accommodates various types of queries. 
For optimal LDP mechanism for SG, one solution is to use split periodic noise cancellation mechanism as per our prior study which provides accuracy in load monitoring and billing. However, due to inherent vulnerability to manipulation attacks, a hybrid user centeric framework that combines multiple privacy preserving models can be used as alternative solution. Such a solution would provide the choice of cost of privacy depending on the SG limitations. For instance, this framework would assign higher cost to LDP based model to the consumers who have batteries installed for energy storage.

\begin{figure}
    \centering
    \includegraphics[width=0.9\linewidth]{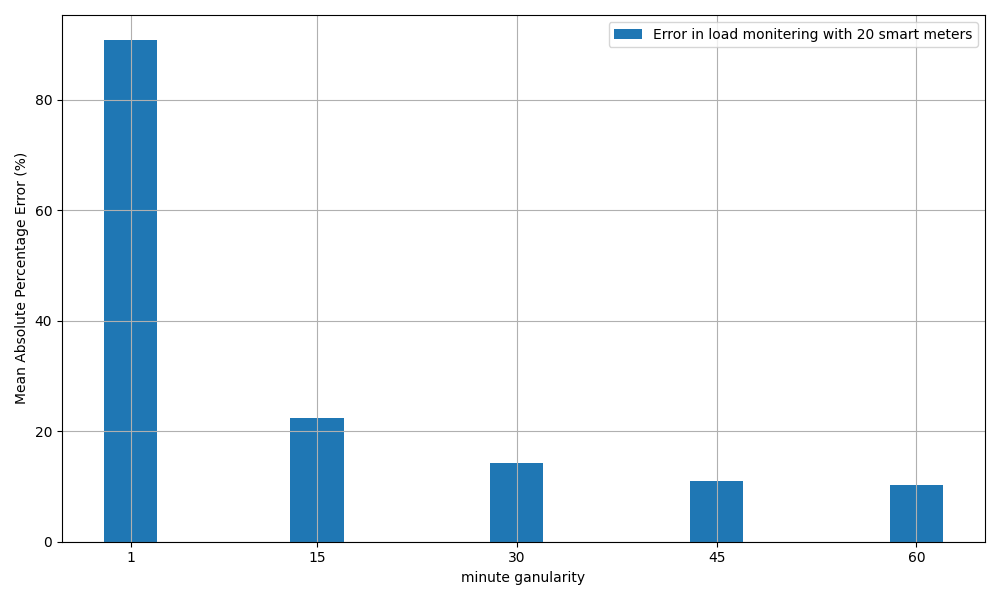}
    \caption{Impact of decreasing granularity in data aggregation for smaller group aggregation.}
    \label{fig:smGranGSize}
    \vspace{-0.5cm}
\end{figure}

\subsection{Optimizing Utility for smaller Data Sets }
The efficacy of LDP needs to be improved for smaller aggregation group size, to guarantee privacy with non-trivial accuracy. Currently, LDP is limited in its applicability to aggregation group sizes of smaller magnitudes, typically in the hundreds \cite{eibl2018influence,SGEnv}. In practical applications for smart grids, such as load monitoring, the number of households or consumers within a given area often falls short of the threshold necessary for LDP to be applicable. Moreover, some tariff policies (e.g.,Time of Use (ToU) billing), require access to the information concerning total energy consumption over shorter intervals, with smaller aggregation group sizes. \par
For optimizing utility for smaller data sets, one solution is to restrict privacy budget for each interval $t$. This can be done by relaxing privacy requirements. This solution will reduce the level of noise added in each interval \cite{edpnct}. Another solution is to reduce the granularity of data while aggregating. For instance if the smart meters are sending LDP protected power consumption data every minute, then load monitoring can be done for every $15$ minutes to decrease the impact of noise on aggregation. The Fig. \ref{fig:smGranGSize} shows the impact of reduced granularity on error in load monitoring with smaller aggregation group size. 
  

\subsection{Protection against Manipulation Attacks}
LDP mechanisms are inherently vulnerable to manipulation attacks where, in the context of smart grids, consumers with the incentive to pay less bills or disrupt the power grid, can hide behind the DP noise and manipulate the input data. Detection and mitigation of such manipulation attacks \cite{deSAMP_fedDP} need further exploration.  \par

 
A mechanism is required that enforces the truthfulness of consumers. One solution is to introduce a cryptographic mechanism to enforce correctness of smart meters such as multiparty computation \cite{verifibe_MPC_DP}. 
Blockchain technology can also be utilized with a consensus mechanism focused on maintaining truthfulness of the data owners. Practicality of such manipulation attack resistant LDP-based models for smart grids is yet to be explored. 

\subsection{Level of Privacy}
Even though the main advantage of an LDP model is the theoretical privacy guarantee, due to the gap in theoretical and experimental LDP-based models for smart grids, there is a lack of understanding of the level of privacy achieved by deploying an LDP-based mechanism. 
LDP is an user-centric approach, hence, it is crucial for the electricity consumers to be aware of how much privacy is achieved by an LDP model, to make informed decisions about the trade-offs between privacy and utility. A clear understanding of level of privacy and privacy parameters empowers consumers to evaluate the level of privacy protection afforded by LDP, enabling them to assess the balance between privacy preservation and the functionality of the system. Hence, a standard and intuitive framework development for LDP-based model for smart grids is much needed. 

\subsection{General LDP mechanism for multiple utility functions}
Most existing literature works focus on LDP aggregation for load monitoring. There is limited work on LDP for services such as billing or load forecasting functions. However, if an LDP model works for one utility function, it does not necessarily gives optimal results for other utility functions. There is a need for a comprehensive study for all the utility functions under a single LDP mechanism. A possible first step is to define the levels of privacy against different privacy attacks, and weigh the cost of each level against all the utility functions. 

\subsection{Hybrid Solutions}
Hybrid solutions, which integrate LDP with other privacy-preserving techniques such as federated learning \cite{deSAMP_fedDP}, homomorphic encryption \cite{dream}, and multiparty computation (MPC) \cite{verifibe_MPC_DP}, have been proposed to address specific limitations of LDP in smart grids. While these models mitigate issues such as utility loss, resistance to collusion, and manipulation attacks, they often introduce new challenges such as high communication and computational overhead. As a result, no existing hybrid solution comprehensively addresses all limitations of LDP, highlighting the need for lightweight models to ensure effective privacy protection in smart grid environments.
\vspace{-0.3cm}
\section{Conclusion}
LDP is being explored as a viable option for preserving privacy of individual consumers' electricity data. This article focuses on two existing LDP-based mechanisms for the smart grids and the limitations of these techniques are discussed in detail. Various factors such as trade-off between data utility vs. privacy, aggregation group size, limited research on setting the privacy budget and sensitivity value, etc., are found to be challenging in adopting LDP-based models for the smart grids. Further, the different utility functions in the context of smart grids and the privacy levels with respect to the privacy attacks need to be explored. Several future directions are presented for the development of a practical privacy preserving LDP-based model for the smart grids.  
\vspace{-0.4cm}
\section*{Acknowledgments}
\small {This publication has emanated from research conducted with the financial support of Taighde Éireann – Research Ireland under Grant number 18/CRT/6222. For the purpose of Open Access, the author has applied a CC BY public copyright licence to any Author Accepted Manuscript version arising from this submission.}

\bibliography{LaplaceLDP}

\begin{thebibliography}{10}
\providecommand{\url}[1]{#1}
\csname url@samestyle\endcsname
\providecommand{\newblock}{\relax}
\providecommand{\bibinfo}[2]{#2}
\providecommand{\BIBentrySTDinterwordspacing}{\spaceskip=0pt\relax}
\providecommand{\BIBentryALTinterwordstretchfactor}{4}
\providecommand{\BIBentryALTinterwordspacing}{\spaceskip=\fontdimen2\font plus
\BIBentryALTinterwordstretchfactor\fontdimen3\font minus
  \fontdimen4\font\relax}
\providecommand{\BIBforeignlanguage}[2]{{%
\expandafter\ifx\csname l@#1\endcsname\relax
\typeout{** WARNING: IEEEtran.bst: No hyphenation pattern has been}%
\typeout{** loaded for the language `#1'. Using the pattern for}%
\typeout{** the default language instead.}%
\else
\language=\csname l@#1\endcsname
\fi
#2}}
\providecommand{\BIBdecl}{\relax}
\BIBdecl

\bibitem{deSAMP_fedDP}
M.~T. Hossain, S.~Islam, S.~Badsha, and H.~Shen, ``Desmp: Differential
  privacy-exploited stealthy model poisoning attacks in federated learning,''
  in \emph{2021 17th International Conference on Mobility, Sensing and
  Networking (MSN)}, Dec. 2021, p. 167–174.

\bibitem{dream}
G.~Ács and C.~Castelluccia, ``I have a dream! (differentially private smart
  metering),'' vol. 6958 LNCS, 2011.

\bibitem{duchi2013local}
J.~Duchi, M.~J. Wainwright, and M.~I. Jordan, ``Local privacy and minimax
  bounds: Sharp rates for probability estimation,'' \emph{Advances in Neural
  Information Processing Systems}, vol.~26, 2013.

\bibitem{LapSGLDP}
Z.~Zheng, T.~Wang, A.~K. Bashir, M.~Alazab, S.~Mumtaz, and X.~Wang, ``A
  decentralized mechanism based on differential privacy for privacy-preserving
  computation in smart grid,'' \emph{IEEE Transactions on Computers}, vol.~71,
  no.~11, pp. 2915--2926, 2022.

\bibitem{Won2016}
J.~{Won}, C.~Y.~T. {Ma}, D.~K.~Y. {Yau}, and N.~S.~V. {Rao}, ``Privacy-assured
  aggregation protocol for smart metering: A proactive fault-tolerant
  approach,'' \emph{IEEE/ACM Transactions on Networking}, vol.~24, no.~3, pp.
  1661--1674, June 2016.

\bibitem{Barbosa2016}
P.~Barbosa, A.~Brito, and H.~Almeida, ``A technique to provide differential
  privacy for appliance usage in smart metering,'' \emph{Information Sciences},
  vol. 370-371, 2016.

\bibitem{eibl2018influence}
G.~Eibl, K.~Bao, P.-W. Grassal, D.~Bernau, and H.~Schmeck, ``The influence of
  differential privacy on short term electric load forecasting,'' \emph{Energy
  Informatics}, vol.~1, no.~1, pp. 93--113, 2018.

\bibitem{SGEnv}
M.~B. Gough, S.~F. Santos, T.~AlSkaif, M.~S. Javadi, R.~Castro, and J.~P.~S.
  Catalão, ``Preserving privacy of smart meter data in a smart grid
  environment,'' \emph{IEEE Transactions on Industrial Informatics}, vol.~18,
  no.~1, pp. 707--718, 2022.

\bibitem{SpectA}
L.~Ou, Z.~Qin, S.~Liao, T.~Li, and D.~Zhang, ``Singular spectrum analysis for
  local differential privacy of classifications in the smart grid,'' \emph{IEEE
  Internet of Things Journal}, vol.~7, no.~6, pp. 5246--5255, 2020.

\bibitem{RR_SG}
N.~Gai, K.~Xue, P.~He, B.~Zhu, J.~Liu, and D.~He, ``An efficient data
  aggregation scheme with local differential privacy in smart grid,'' in
  \emph{2020 16th International Conference on Mobility, Sensing and Networking
  (MSN)}, 2020, pp. 73--80.

\bibitem{prefLimitationLap}
S.~Wang, X.~Luo, Y.~Qian, J.~Du, W.~Lin, and W.~Yang, ``Analyzing preference
  data with local privacy: Optimal utility and enhanced robustness,''
  \emph{IEEE Transactions on Knowledge and Data Engineering}, vol.~35, no.~8,
  pp. 7753--7767, 2023.

\bibitem{edpnct}
K.~Hafeez, D.~O'Shea, T.~Newe, and M.~H. Rehmani, ``E-dpnct: an enhanced attack
  resilient differential privacy model for smart grids using split noise
  cancellation,'' \emph{Scientific Reports}, vol.~13, no.~1, p. 19546, 2023.

\bibitem{LapvsPWvsH}
N.~Wang, X.~Xiao, Y.~Yang, J.~Zhao, S.~C. Hui, H.~Shin, J.~Shin, and G.~Yu,
  ``Collecting and analyzing multidimensional data with local differential
  privacy,'' in \emph{IEEE 35th International Conference on Data Engineering
  (ICDE)}, 2019, pp. 638--649.

\bibitem{d-priv}
N.~Fernandes, A.~McIver, C.~Palamidessi, and M.~Ding, ``Universal optimality
  and robust utility bounds for metric differential privacy,'' in \emph{2022
  IEEE 35th Computer Security Foundations Symposium (CSF)}, 2022, pp. 348--363.

\bibitem{verifibe_MPC_DP}
F.~Kato, Y.~Cao, and M.~Yoshikawa, ``Preventing manipulation attack in local
  differential privacy using verifiable randomization mechanism,'' in
  \emph{Data and Applications Security and Privacy XXXV}, K.~Barker and
  K.~Ghazinour, Eds.\hskip 1em plus 0.5em minus 0.4em\relax Cham: Springer
  International Publishing, 2021, pp. 43--60.

\end{thebibliography}
\bibliographystyle{IEEEtran}

\vspace{-1cm}

\begin{IEEEbiographynophoto}{Khadija Hafeez}
received her masters in computer science and software engineering degree from Lahore University of Management Sciences (LUMS) and National University of Science and Technology (NUST) respectively. She is currently a Ph.D. student at Munster Technological University (MTU), Ireland. Her research interest are data privacy, smart grids, IoT and computational thinking. 
\end{IEEEbiographynophoto}
\vspace{-1cm}
\begin{IEEEbiographynophoto}{Mubashir Husain Rehmani}
 (M’14-SM’15, SFHEA) is working as Lecturer with the Department of Computer Science, Munster Technological University (MTU), Cork, Ireland. He received his Ph.D. from the University Pierre and Marie Curie, Paris, in 2011. He currently serves as an Area Editor of the IEEE Communications Surveys and Tutorials and IEEE Open Journal of Communications Society. He also serves as an Associate Editor for IEEE Transactions on Green Communication and Networking and IEEE Transactions on Cognitive Communications and Networking. He is the recipient of Highly Cited Researcher™ award thrice in 2020, 2021, and 2022 by Clarivate, USA. His performance in this context features in the TOP $1\%$ by citations in the field of Computer Science and Cross Field in the Web of Science™ citation index. In Oct 2022, he received Science Foundation Ireland’s CONNECT Centre’s Education and Public Engagement (EPE) Award 2022. Contact him at mshrehmani@gmail.com.
\end{IEEEbiographynophoto}
\vspace{-1cm}
\begin{IEEEbiographynophoto}{Sumita Mishra}
is a Professor and the Graduate Program Director of Cybersecurity at Rochester Institute of Technology, Rochester, NY, USA. She earned her doctoral degree in Electrical Engineering at the University at Buffalo in 2001 and has been a faculty member at RIT for 18 years. Her research interests are in applied cryptography, critical infrastructure protection, and cybersecurity pedagogy. Funding sources for Dr. Mishra’s current and past projects include the National Science Foundation, the Department of Homeland Security, and Air Force Research Labs. She serves on the editorial board of Elsevier Computer Communications and several Program Committees of IEEE and ACM conferences. She has published over 75 articles in reputed journals and academic conferences in her field.
\end{IEEEbiographynophoto}
\vspace{-1cm}
\begin{IEEEbiographynophoto}{Donna O'Shea}
(PhD-07,MEng-04,BSc-02) is a leading national figure in cybersecurity research/innovation. She holds the position of Chair in Cybersecurity at Munster Technological University (MTU). She has extensive experience in securing funding and leading large-scale national cybersecurity initiatives including the ‘Cyber-Ireland’ Industry Cluster/EI Innovator’s Initiative ‘Cyber-Innovate’ focused in increasing the number of SMEs/HPSUs in cybersecurity/HEA-HCI3 ‘Cyber-Skills’ project on responding to the critical shortage of cybersecurity professionals and Research Ireland’s ‘Cyber-Futures’ EPE project in cybersecurity. She is involved with government and the National Cyber Security Centre(NCSC) on policy development and collaborates extensively with industry. She is an Funded Investigator, in the CONNECT Research Ireland Centre for Future Networks with a strong track record in doctoral supervision. Previously, she was Head of Department and Lecturer in MTU Computer Science(2013-2020) and worked in IBM as a Software Developer(2008-2013). 
\end{IEEEbiographynophoto}
\end{document}